\documentclass{jfm}
\usepackage{graphicx}
\usepackage{epstopdf, epsfig}
\usepackage{natbib}
\usepackage{subfigure}
\usepackage{amssymb}
\usepackage{amsbsy}
\usepackage{color}
\usepackage{ulem}

\newcommand{\rr}[1]{{\textcolor{blue}{#1}}}

\newcommand{\DEL}[1]{\textcolor{green}{\sout{#1}}}
\newcommand{\REM}[1]{{}}
\shorttitle{Surface tension effects on a vortex at the fluid interface.}
\shortauthor{R. Ramadugu, P. Perlekar and R. Govindarajan}

\title{Surface tension as the destabiliser of a vortical interface}

\author{Rashmi Ramadugu\aff{1}
  Prasad Perlekar \aff{1}
   \corresp{\email{perlekar@tifrh.res.in}},
 \and Rama Govindarajan\aff{2}}

\affiliation{\aff{1}TIFR Center for Interdisciplinary Sciences, Tata Institute of Fundamental Research, 500046, Gopanpally, Hyderabad, India
\aff{2}International Centre for Theoretical Sciences, Tata Institute of Fundamental Research, Shivakote,
Bengaluru 560089, India}

\begin{document}

\maketitle

\begin{abstract}
We study the dynamics of an initially flat interface between two immiscible fluids, with a vortex situated on it. We show how surface tension causes vorticity generation at a general curved interface. This  creates a velocity jump across the interface which increases quadratically in time, and causes the Kelvin-Helmholtz instability. Surface tension thus acts as a destabiliser by vorticity creation, winning over its own tendency to stabilize by smoothing out interfacial perturbations to reduce surface energy. We further show that this instability is manifested within the vortex core at times larger than $\sim (k We)^{1/4}$ for a Weber number $We$ and perturbation wavenumber $k$, destroying the flow structure. The vorticity peels off into small-scale structures away from the interface. Using energy balance we provide the growth with time in total interface length. A density difference between the fluids produces additional instabilities outside the vortex core due to centrifugal effects. We demonstrate the importance of this mechanism in  two-dimensional turbulence simulations with a prescribed initial interface. 

\end{abstract}
\maketitle
\section{Introduction}

The interaction between a vortex and an interface may be considered a building block in the turbulent flow of immiscible fluids. We study this building block and show that it is prone to a Kelvin-Helmholtz (KH)  instability {\it created} by surface tension. We will distinguish our flow from KH instabilities at immiscible interfaces across which a velocity jump is externally imposed. This latter class of problems has been well studied, and we begin by discussing a few studies. The effect of surface tension on the primary KH roll-up process was studied using two-dimensional numerical simulations by \cite{, fakhari2013multiple}. The main finding was that surface tension has a stabilising effect on the flow.  \cite{hou1997long} showed that KH roll ups form only if surface tension is low, and the range of unstable scales diminish with increase in surface tension.  \cite{rangel1988nonlinear} showed for a KH instability that non-zero surface tension results in an increase of the stable regime. \cite{tauber2002nonlinear} investigated KH instability in density matched fluids at large Reynolds numbers. They find that the non-linear roll-up at low surface tension is similar to that at zero surface tension and high surface tension results in a nearly flat interface with no roll up.
Consistently across these studies, surface tension thus acts as a stabiliser, by effecting a reduction in interface length and thus suppressing KH roll-up. In fact, in a vast variety of flow situations, surface tension suppresses large wave number perturbations, thereby decreasing interface area.

It has been long known that the same quality of bringing about a reduction in surface area can make surface tension a destabilising agent, but in other contexts.  Famously, \citep{strutt1878instability} surface tension can destabilise liquid jets and break them up into droplets by the so-called Plateau-Rayleigh instability. Again, this happens because of the propensity of higher surface tension to effect a reduction in surface area in a circular flow geometry. In a planar geometry,  \cite{biancofiore2017} studied two parallel interfaces separating three immiscible density-matched fluids with linear shear profiles in a Taylor-Caulfield configuration. This system, which is stable without surface tension, is shown to display an instability when there is a phase lock between counter-propagating capillary waves. Thus, wave interactions cause surface tension to act as destabiliser. By a similar mechanism, surface tension at the interface in planar jets and wakes at high enough levels of shear can produce global instabilities \citep{tammisola2012surface}. 

Surface tension has occasionally been reported as giving rise to small-scale structures. \cite{zhang2001surface} again studied the effect of surface tension on the KH instability. At high surface tension, they showed the generation of small-scale vortices in the late stages of evolution, giving rise to a positive contribution of surface tension to flow enstrophy despite a negative contribution to kinetic energy. 
 The recent study of \citep{tavares2020immiscible} shows evidence, in Rayleigh Taylor turbulence, of a greater preponderance of smaller scales in immiscible flows with surface tension as compared to miscible flows.  

We propose here a new mechanism for the destabilising action of surface tension. The dynamics due to a single vortex placed at an interface between two fluids is studied in the absence of gravity and viscosity. This geometry is similar to \cite{dixit2010}, but that study was at zero surface tension. We show analytically how vorticity is produced by surface tension at the interface, and how this makes the flow unstable.  Our direct numerical simulations (DNS) confirm our analytical predictions, and show the differences in evolution of vorticity and the interface. 

\section{Problem description}\label{sec:statement}
Two immiscible fluids of constant densities $\rho_0$ and $\rho_1$ lie on either side of an initially flat interface in a two-dimensional system. The fluids are incompressible and inviscid, and the continuity and momentum equations they each satisfy are
\begin{eqnarray}
\frac{D \rho}{D t} &=& 0, \qquad  
\pmb{\nabla} \cdot \mathbf{u} = 0, \label{cont}  \\
\rho  \frac{D \mathbf{u}}{D t} &=& -\pmb{\nabla} P + \mathbf{F}_{\sigma},  \label{nav-stks0}
\end{eqnarray}
where $P$ is the pressure, $\mathbf{u} = \{u_{r},u_{\theta}\} $ are the radial and azimuthal velocities respectively,
\begin{equation}
\mathbf{F}_{\sigma} = \sigma \kappa \delta(\mathbf{x}-\mathbf{x}_s)\mathbf{n}
\label{surften}
\end{equation} 
is the surface tension force density, $\sigma$ is the surface tension, $\kappa$ is the curvature, $\mathbf{x}=\{r,\theta\}$, the subscript $s$ stands for a location on the interface, $\delta(.)$ is the Dirac delta function and $\mathbf{n}$ is the unit normal to the interface at $\mathbf{x}_s$. 
\begin{figure}
\includegraphics[scale=0.31]{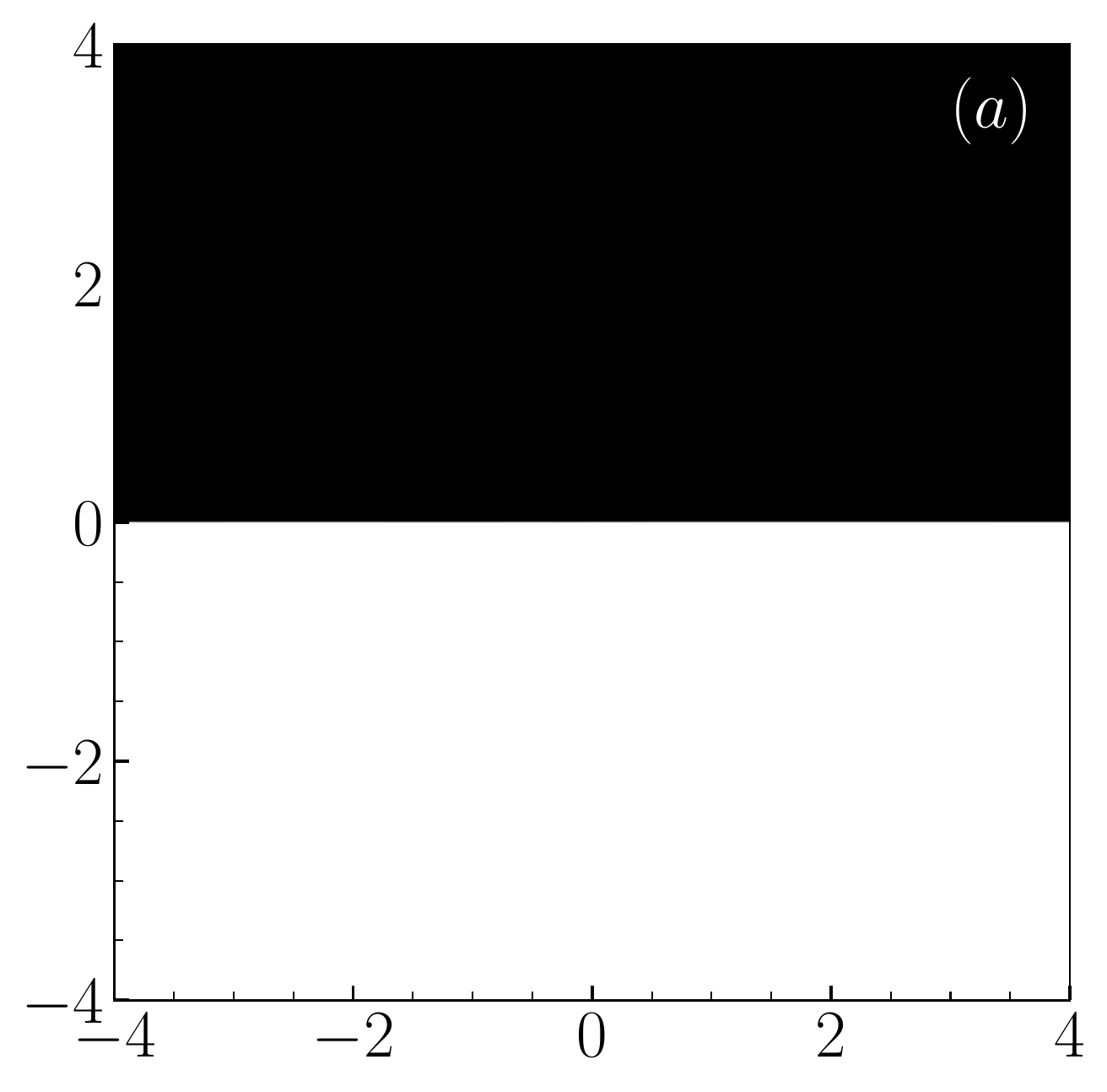}
\includegraphics[scale=0.31]{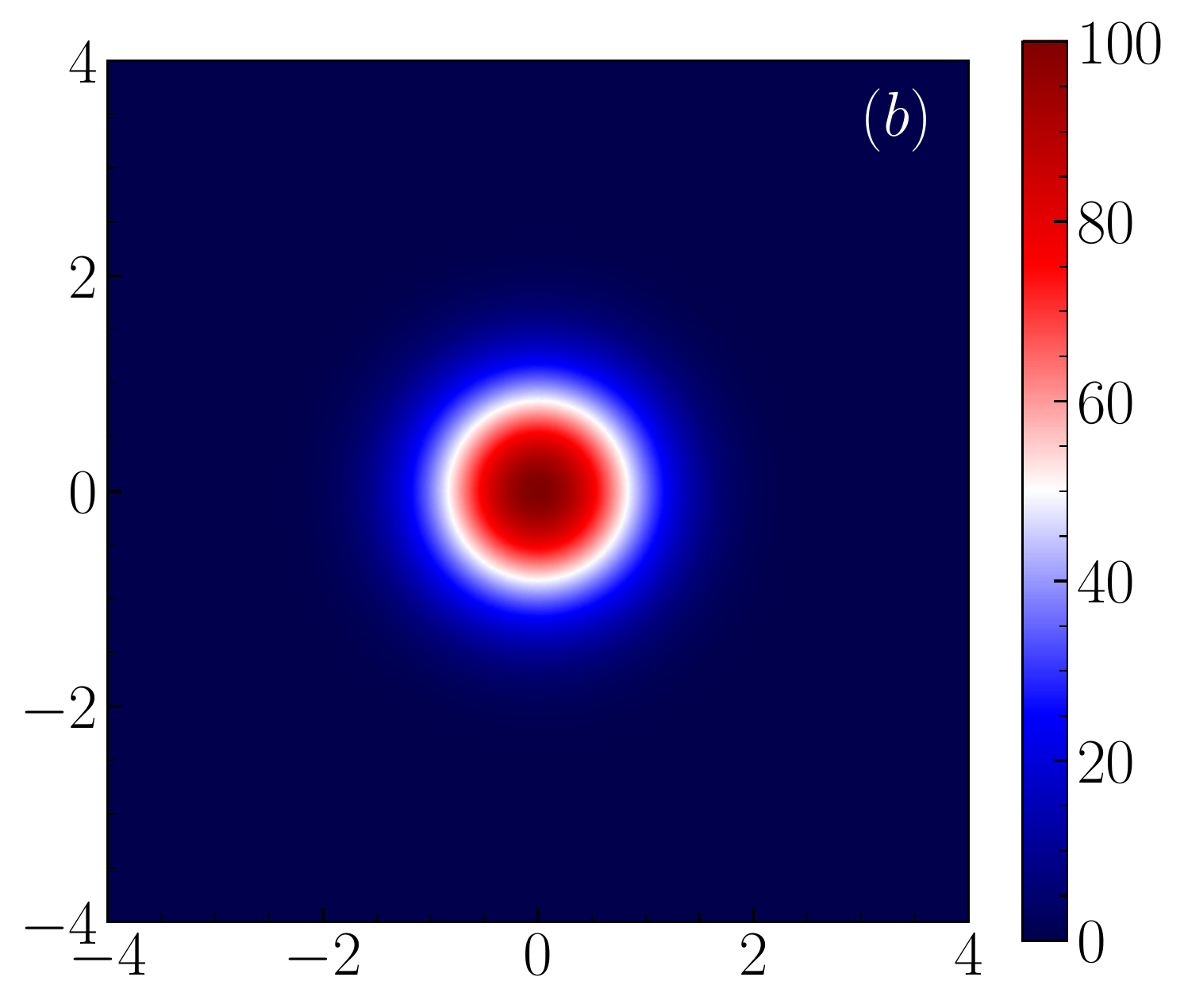}
 \includegraphics[scale=0.31]{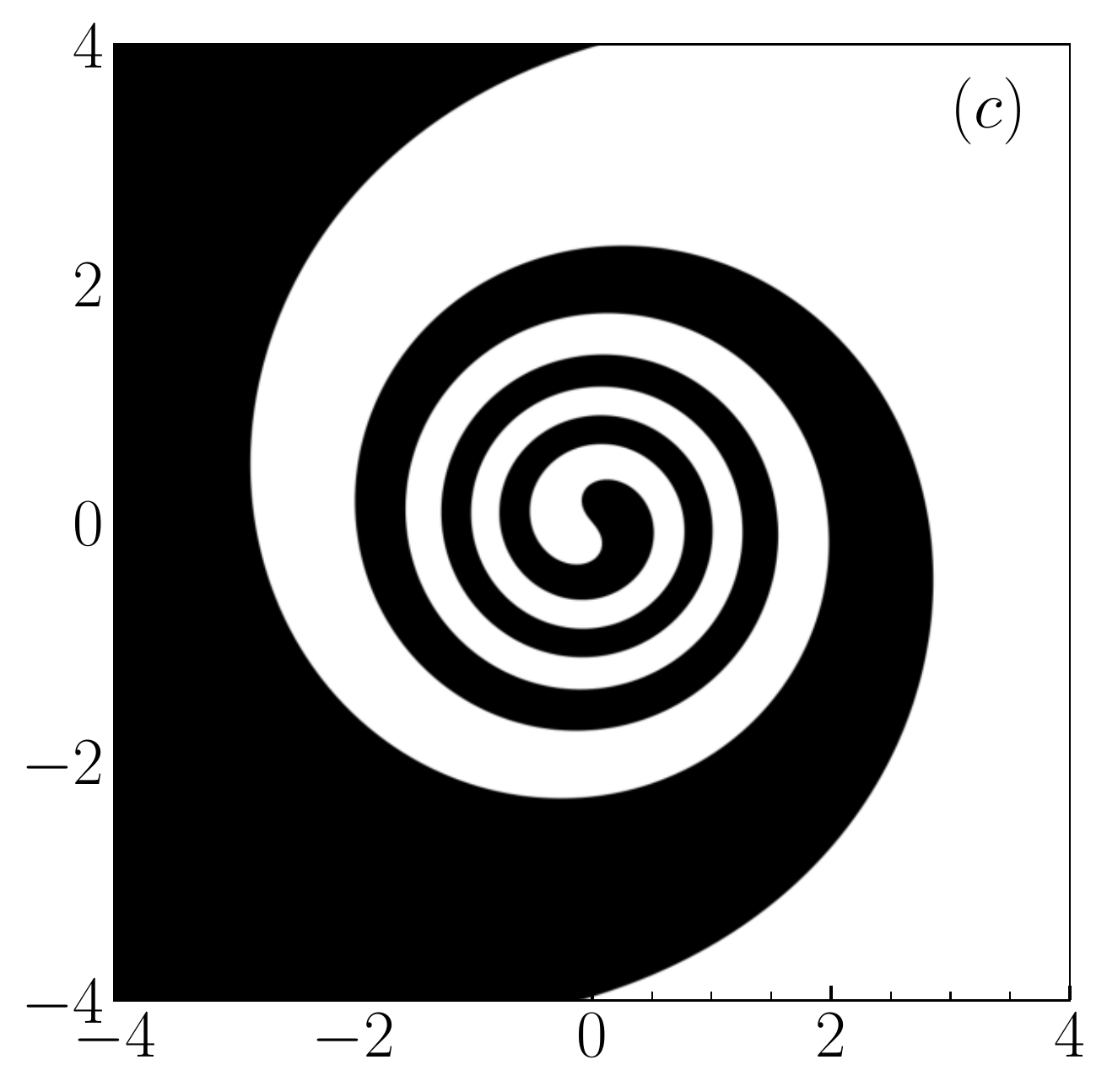}
\caption{ (a)  The initial density field with an equal volume of two fluids shown in black and white.(b) The initial vorticity field, (Lamb-Oseen vortex with core radius($r_c$) = 0.1).  (c) The fluid field at a later time, when the two fluids have the same density and surface tension is vanishingly small.}
\label{Fig:density_profile}
\end{figure}
The interface passes through the origin. A Lamb-Oseen vortex of circulation $\Gamma$ and core radius $r_c$ is placed with its centre at the origin at time $t=0$, as shown in Fig. \ref{Fig:density_profile}(a,b). When surface tension is zero, and the two fluids have identical densities, each fluid particle moves strictly in a circular path, with an azimuthal velocity given by 
\begin{equation}
{U}  = \frac{\Gamma}{2 \pi r} \left[1- \exp\left(-q\right)\right], \label{azivel}
\end{equation}
where for ease of algebra we have defined $q \equiv (r/r_c)^2$. The total angle,  $\theta_s$, swept out up to time $t$ by the interface at $r_s$, is a linearly increasing function of time given by
  \begin{equation}
\theta_s = \frac{\Gamma t}{2 \pi r_s^2} \left[1- \exp\left\{-q_s\right\}\right]. 
\label{inter_eq_exp}
\end{equation}
For $r_s  \gg r_c$, Eq. \ref{azivel} reduces to $ {U} = \Gamma/(2 \pi r_s)$ for a point vortex, and  an initially flat interface will wind up into an ever-tightening spiral \citep{dixit2010}, given by $r_s^2 \theta_s = \Gamma t$. Thus, away from the vortex core, at every instance of time, the interface describes a different Lituus spiral, which is one among the Archimedean class of spirals, as seen at a typical time in Fig. \ref{Fig:density_profile}(c). 
 
The relevant non-dimensional numbers are the Weber number ($We$) which is a ratio of the inertial force to the surface tension force and the Atwood number ($At$)  which is a measure of the density contrast between the two fluids:
 \begin{eqnarray}
 We \equiv \frac{ \overline{\rho} U_c^2 r_c}{\sigma}, \quad \quad At \equiv \frac{\Delta \rho}{\rho_0 + \rho_1}, 
\end{eqnarray}
where  $\Delta \rho = \rho_0 - \rho_1$, $\overline{\rho} = (\rho_0 + \rho_1)/2$, and $U_c=\Gamma/(2 \pi r_c) [1- \exp(-1)]$ is the azimuthal velocity of the fluid at $r_c$.
It is sometimes convenient to choose the radial distance $r$ from the vortex centre as our length scale, so the Weber number $We_r$ thus defined is a local quantity. At a given radius, the inertial time-scale $2\pi r^2/\Gamma$, for the wind-up of the spiral, is shorter than the time scale $T_\sigma=\sqrt{\rho r^3/\sigma}$, at which surface tension effects will be visible in the basic flow, by a factor $We_r^{1/2}$. For large $We_r$ and in the absence of instabilities, we may neglect the effect of surface tension on the interface shape up to a non-dimensional time of $We_r^{1/2}$ and the base velocity given by the Lamb-Oseen vortex will dictate the interface shape. 

 \section{Vorticity generation on the interface and the Kelvin-Helmholtz instability}\label{Sec:vorticity}
An important aspect of this dynamics is the creation of vorticity at the interface by the surface tension and by the density contrast (baroclinic torque). This may be seen with Eq. \ref{nav-stks0} rewritten in the vorticity formulation as
\begin{equation}\label{Eq:vort1}
 \frac{D \Omega}{Dt} =  \frac{D (\Omega_\sigma + \Omega_b)}{Dt} = \frac{1}{\rho} \pmb{\nabla} \times \mathbf{F}_{\sigma} - \frac{1}{\rho^2}\pmb{\nabla} \rho \times \pmb{\nabla}P,
\end{equation}
where $\Omega=\pmb{\nabla} \times \mathbf{u}$ is the vorticity, here pointing in the  out-of-plane direction.   $\Omega$ has contributions from the surface tension, $\Omega_{\sigma}$, and buoyancy, $\Omega_b$, represented respectively by the first and second terms on the right hand side. An examination of Eqns. \ref{surften} and \ref{azivel} makes it clear that no vorticity will be generated by a perfectly flat interface or a perfectly circular one. But when the shape of the interface deviates from these geometries, vorticity may be generated by both density gradients and surface tension. Our focus here is on surface tension as a generator of vorticity, so we discuss the case of $At=0$ below.


Integrating the first term in Eq. \ref{Eq:vort1} at a given radial location, we obtain a time-dependent vorticity at the interface as follows. Consider $f(r,\theta) = r-r_s(\theta)$ which vanishes on the interface, with a  normal $\mathbf{n} \equiv \pmb{\nabla}f/|\pmb{\nabla} f| $, and curvature, $\kappa \equiv -\pmb{\nabla} \cdot  \mathbf{n}$. We have
 \begin{equation}
\pmb{\nabla} \times \mathbf{F}_{\sigma} =-\frac{1}{r}\partial_{\theta} F_{\sigma,r} = \frac{\sigma}{r} \partial_{\theta}\bigg(\frac{r_s^3 + 2r_s (\partial_{\theta} r_s)^2 - r_s^2 \partial_{\theta \theta}r_s}{[r_s^2 + (\partial_{\theta} r_s)^2]^2}  \bigg),  \label{Eq:curlf2}
\end{equation}
which leads, upon considerable simplification, to
\begin{eqnarray}
\frac{r_c}{U_c}	\Omega_{\sigma} &=& \frac{(1-e^{-1}) e^q   }{4 We \beta^3} \bigg\{ 8 (q^2 + \beta)^2 \left[1-\frac{1}{\chi^2}\right] - 2 \bigg[4 q^4 + 11 q^3 + 18 q^2 +  10 q  + 5 +
	\nonumber \\
	&&  (2 q^3 - 13 q^2 -10 q  -  10) e^q   + 5 e^{2q}\bigg]  \left[1-\frac{1}{  \chi}\right] - \beta^2 \log \chi\bigg\} \delta\left(\frac{r}{r_c}\theta - \frac{r}{r_c} \theta_s\right), \nonumber \\
& \equiv & \frac{\Delta U}{U_c}
\delta\left(\frac{r}{r_c}\theta - \frac{r}{r_c} \theta_s\right)
	\label{Eqn:vort_new},
 \end{eqnarray}
 where  $\beta= [q + 1 - e^q]$, $\chi =  1+ \left[ 2t_n  \beta/ (q e^q)\right]^2$,  $t_n  =  t/T_c$, and $T_c = 2\pi r_c^2/\Gamma$ is the inertial time scale at $r_c$. Due to the vorticity created at the interface, the two fluids on either side move with different velocities. We denote the velocity components parallel to the interface on either side by $U_{||}$ and $U_{||} + \Delta U$, with the  jump $\Delta U$ across the interface  given by Eq. \ref{Eqn:vort_new}. It follows that the interface must be subject to the Kelvin-Helmholtz (KH) instability. For the complete problem, an analytical dispersion relation is not possible to write down, but approximate estimates of the instability growth rates may be written down in two limiting regimes.
 
 Well within the core, we have $r\ll r_c$, so $q \ll 1$. Taylor expanding in this limit, after some algebra, and retaining the first surviving term in the expansion, Eq. \ref{Eqn:vort_new} reduces to
 \begin{equation}
\frac{\Delta U}{U_c} = -\frac{9 (1-e^{-1}) t_n^2 }{ 2 We }.
\label{Eqn:vort_smallr}
 \end{equation}
The vorticity produced, and thus the jump in velocity, are quadratic in time and proportional to the surface tension. 
In this limit, the interface can be closely approximated by a straight line rotating at a constant rate.  We may then write the relevant dispersion relation  \citep{chandrasekhar1981hydrodynamic}, in the case where there is no density contrast ($At=0$),  in non-dimensional form as
\begin{equation}
\frac{\omega}{k {U}_c} = -\frac{\Delta U} {2 U_c}  \pm \left[\frac{k r_c}{2 We} -  \frac{1}{4}\left\{\frac{\Delta U}{U_c}\right\}^2 \right]^{1/2},
\label{Eqn:omeg2_new}
\end{equation}
where the real and imaginary parts of $\omega$ respectively give the frequency and growth rate of a perturbation of wavenumber $k$.
The first term within the square bracket stands for the standard stabilising action of surface tension, increasing with wavenumber. The second term on the other hand indicates the destabilising effect of surface tension. As seen  from Eq. \ref{Eqn:vort_smallr}, it is quadratic in the surface tension, while the stabilising term is only linear in this quantity. Moreover, we know from Eq. \ref{Eqn:vort_smallr} that $\Delta U$ increases quadratically in time, so the destabilising action of surface tension must win over its stabilising action at some time for any Weber number. In other words, the interface within the core becomes KH unstable when 
\begin{equation}
   t_n > \frac{(8 k r_c We)^{1/4}}{3 (1-e^{-1})^{1/2}}.
   \label{Eqn:tneq}
\end{equation}
 The higher the surface tension, the faster the instability grows. But at any given time, for any finite $We$, there is a cutoff wavenumber beyond which flow is stable. 

On the other hand, well outside the vortex core, $q \gg 1$, and to leading order in $1/q$ we have
\begin{equation}
\frac{\Delta U}{U_c} = -\frac{40 (1-e^{-1}) t_n^2 }{q^2 We }, \label{Eqn:vort_bigr}
 \end{equation}
 so the vorticity on the interface decreases as $r^{-4}$. For estimating the instability, we may, by following a procedure analogous to \cite{dixit2010}, approximate the interface as a circle at a radius $r$\DEL{,} and perturbing it at azimuthal wavenumber $m$, to get
\begin{eqnarray}
\frac{\omega r_1}{mU_c} =  q^{-1/2} + \bigg( \frac{1-At}{2}\bigg)\left( \frac{\Delta U}{U_c}\right)  && \pm q^{-1/2} \bigg[   \bigg\{\frac{(1-At) q}{2m} -  \bigg( \frac{1-At^2}{4}\bigg)\bigg\} \bigg(\frac{\Delta U}{U_c}\bigg)^2 +  \nonumber \\ && - \frac{(1-At) q^{1/2}}{m} \frac{\Delta U}{U_c}  -\frac{At}{m}  + \frac{m}{2 We} \bigg]^{1/2}.
\label{Eqn:omeg1_new}
\end{eqnarray}
 On examining this expression in the light of Eq. \ref{Eqn:vort_bigr}, we see that at $At=0$ the largest destabilising term is $O(q^{-3/2})$ smaller than the stabilizing term, so  instability is not expected. When $At>0$ however, the spiralling interface can be unstable, as found by \cite{dixit2010}, but now surface tension stabilizes the flow at high azimuthal wavenumber. Thus, the instability within a vortex core is driven by surface tension and that outside  by density differences.

\section{Direct Numerical Simulations (DNS)}

\subsection{Simulation details}

We conduct DNS using an open-source Volume of Fluid (VOF) code Basilisk to solve Eqs. \ref{cont} and Eq. \ref{nav-stks0} \citep{popinet2018}.  
We place a Lamb-Oseen vortex at the interface in the center of the domain as shown in Fig. \ref{Fig:density_profile} (a,b) and  allow it to evolve with time. Since our computational domain, of length $L=5\pi r_c$, is much larger than our vortex, the far-field boundary conditions do not affect the results. We use free-slip conditions on all sides in our inviscid flow, and have checked that periodic boundary conditions give practically indistinguishable answers. Note that our simulations include non-Boussinesq effects when $At \ne 0$. We conduct  DNS in a square box of length $2\pi$, discretize it with $2048^2$ collocation points, and vary the 
Weber numbers $(We): 10040, 100.4, 33.3,$ and $12$.
We restrict our simulations to times over which the sum of the interfacial and kinetic energies is constant.


 \subsection{Vorticity generation and the resulting instabilities}\label{Sec:vorticity_DNS}

\begin{figure}
    \includegraphics[scale=0.26]{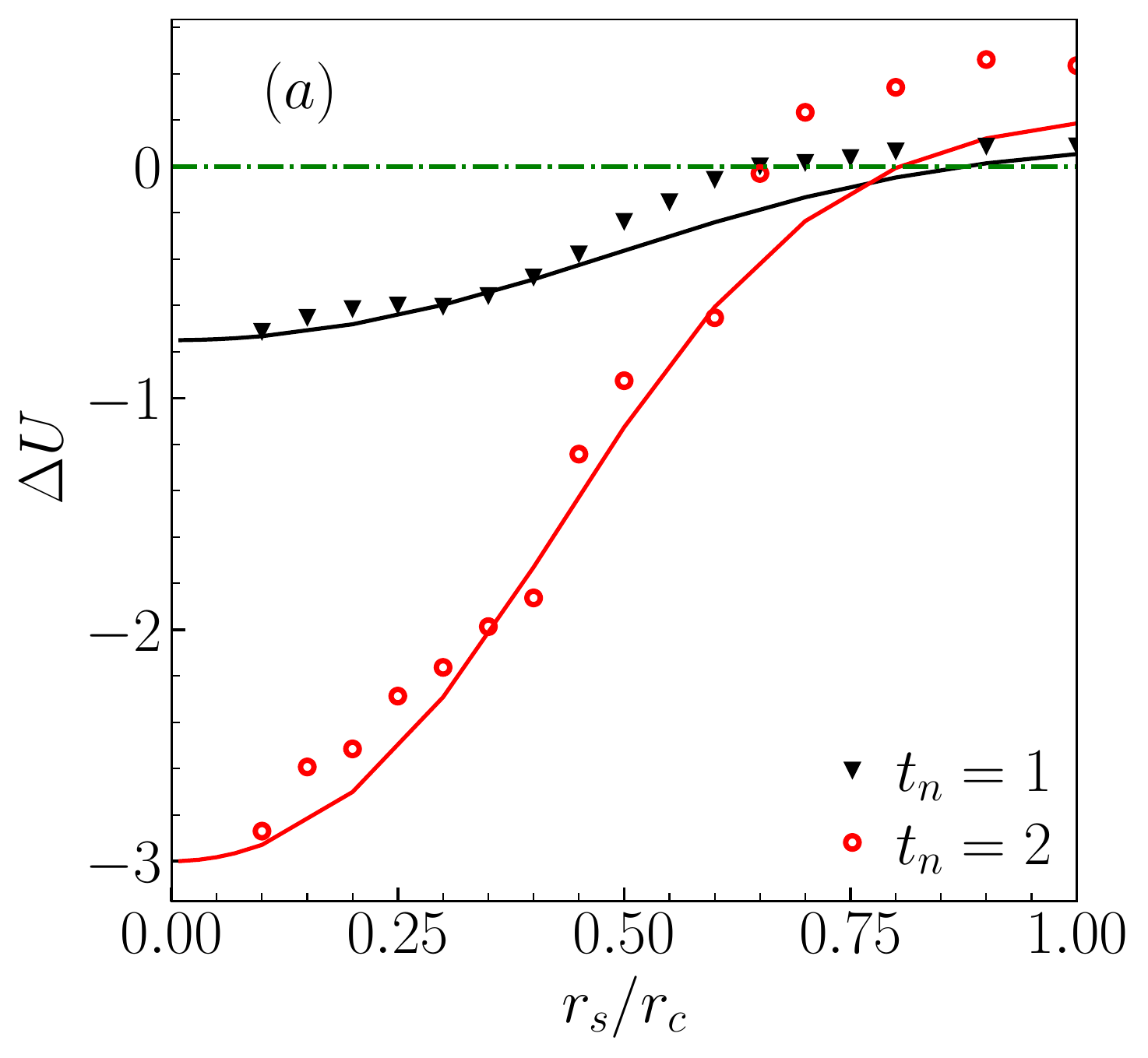}
    \includegraphics[scale=0.28]{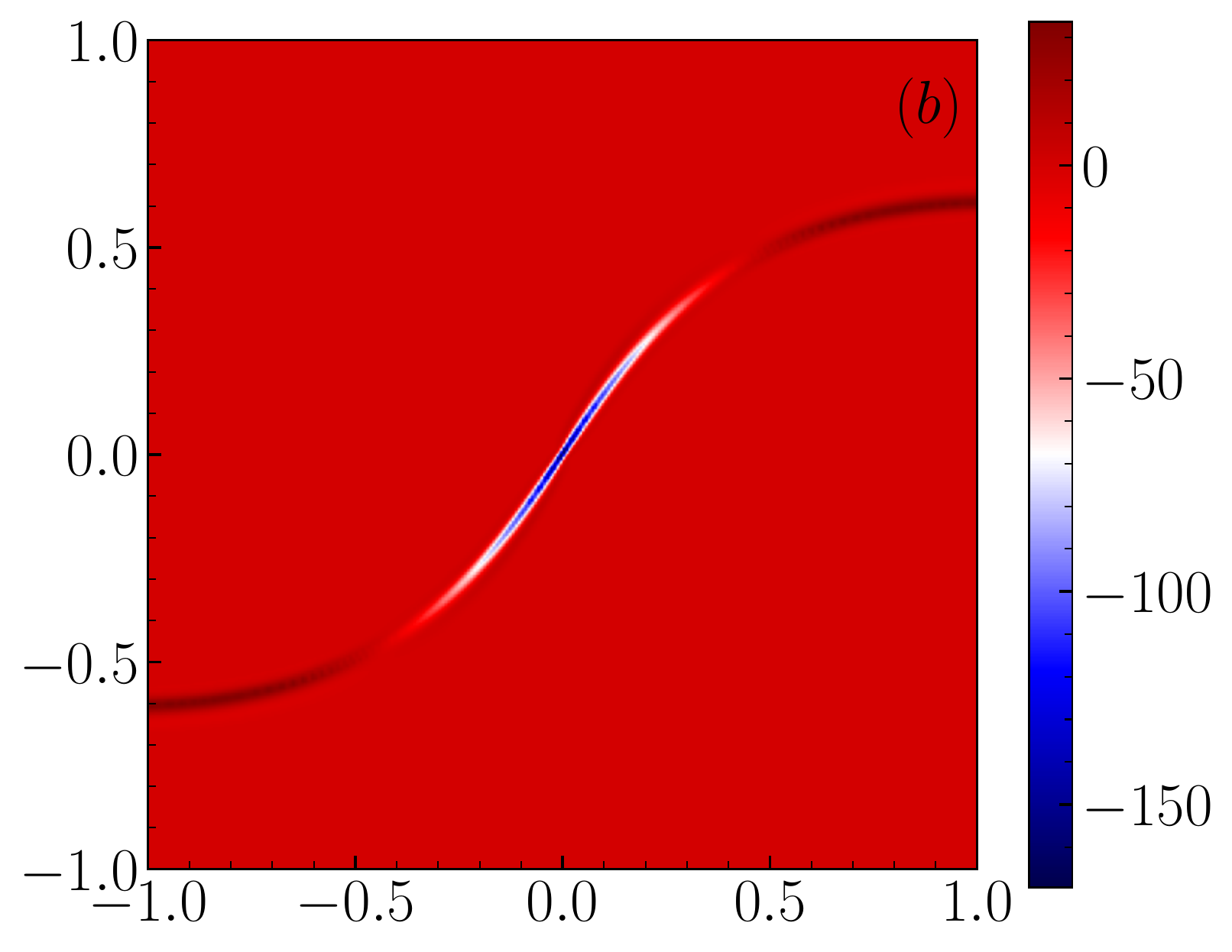}
    \includegraphics[scale=0.28]{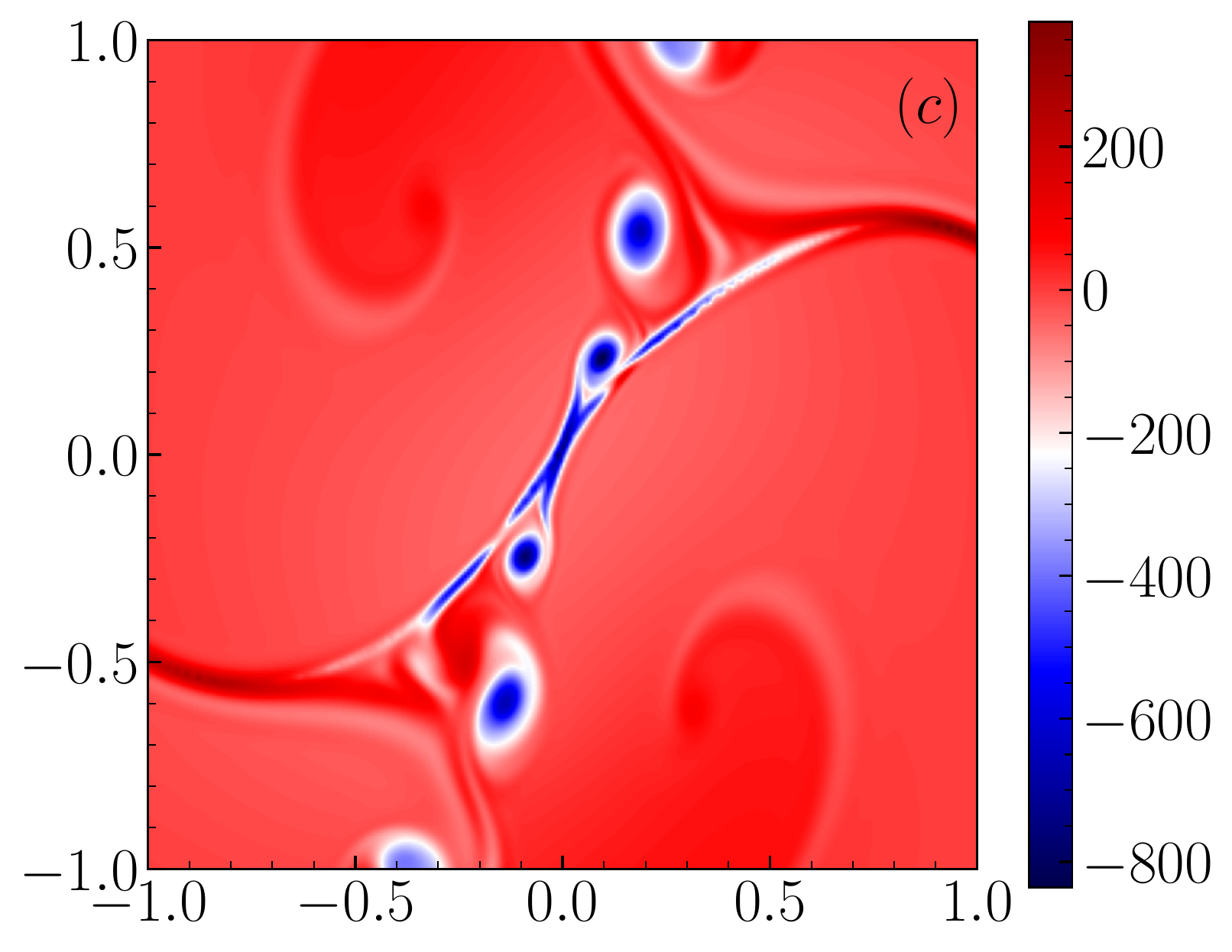}
    \caption{Vorticity generation with $We=12$. (a) Velocity jump measured across the interface, as a function of interface location within the vortex core. Symbols: DNS, line: Eq. \ref{Eqn:vort_new}. The perturbation vorticity from the DNS is shown for time $t_n=1$ in (b) and for $t_n = 15$ in (c). The instability is well-developed by $t_n=15.$ All lengths are scaled by $r_c$. }
   
    \label{fig:vort_compare}
\end{figure}
Up to time $t_n \sim We^{1/2}$ we expect the vorticity generated on the interface in the numerical simulations to closely follow Eq. \ref{Eqn:vort_new}.  This is shown to be the case in Fig. \ref{fig:vort_compare}(a), where $t_n=1$ and $t_n = 2$  for $We=12$. To calculate the velocity jump across  the interface at the point $r_s$,  we numerically integrate the vorticity along the normal. We have checked that $\Delta U$ is insensitive to further grid resolution. As predicted, $\Delta U \sim t_n^2$. Interestingly, the velocity jump changes sign, going from negative to positive below $r_c$. There is one more sign change at large $r$ (not shown here), and the velocity jump in the distant spiral arms is negative again, though very small. Fig. \ref{fig:vort_compare}(b) and (c) show the perturbation vorticity at two times,  calculated by subtracting the initial vorticity from the instantaneous field. It is seen that at the larger time, instability has set in, which is consistent with the prediction of instability when $t_n > 2.08$ for $kr_c = 2 \pi$ by the Eq. \ref{Eqn:vort_smallr}. The instability becomes visible in the simulations to the naked eye at around $t_n=5$ (see movies in the supplementary material). For the case where surface tension ($We = 10040$) is much lower, the time above which instability can occur, from Eq. \ref{Eqn:vort_smallr}, is very similar to that for $We=12$, but because the growth rate is minuscule as predicted by the Eq.  \ref{Eqn:omeg2_new} , an instability does not become visible during the time of our simulation.

\begin{figure}
\begin{center}
\includegraphics[scale=0.275]{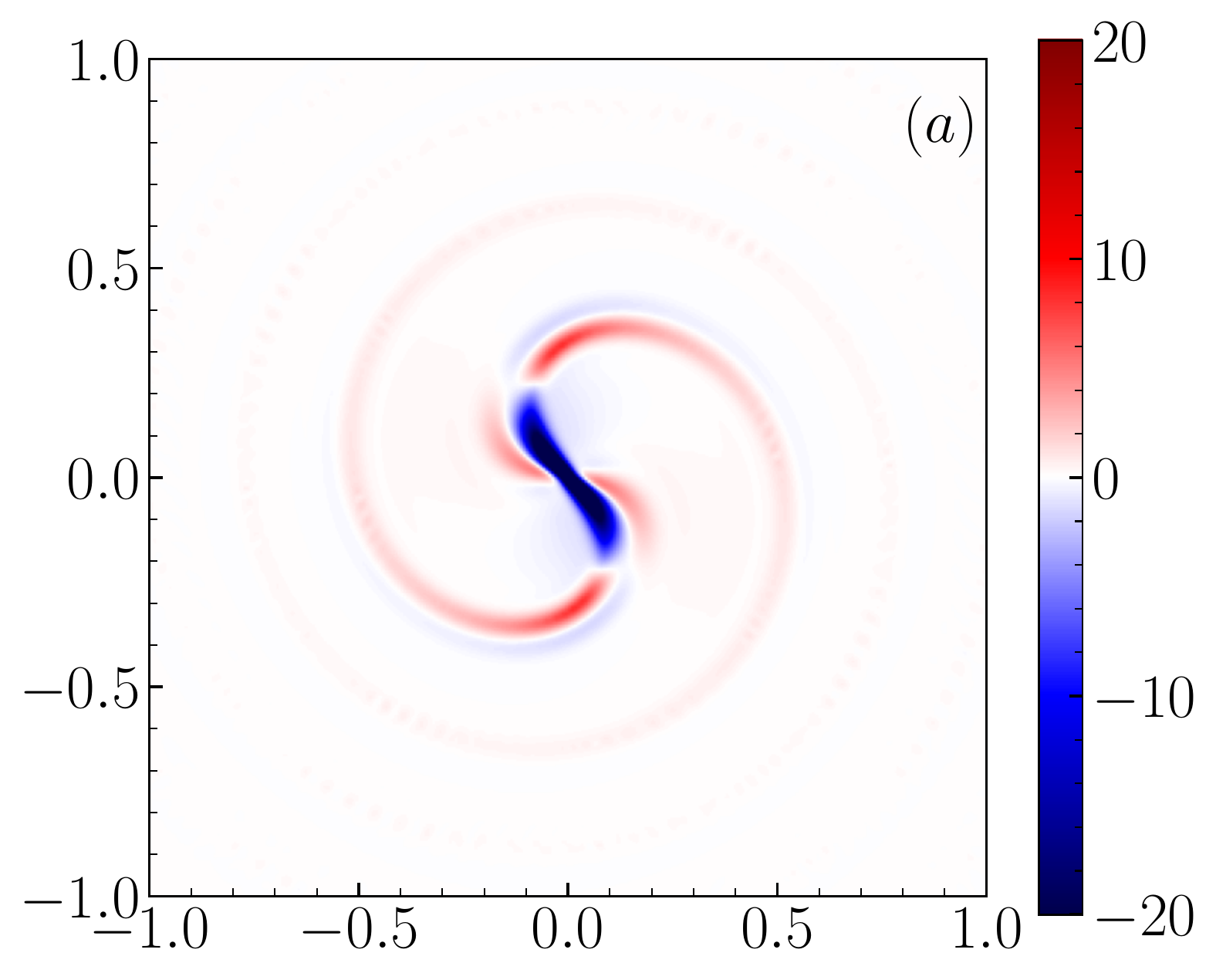}
\includegraphics[scale=0.275]{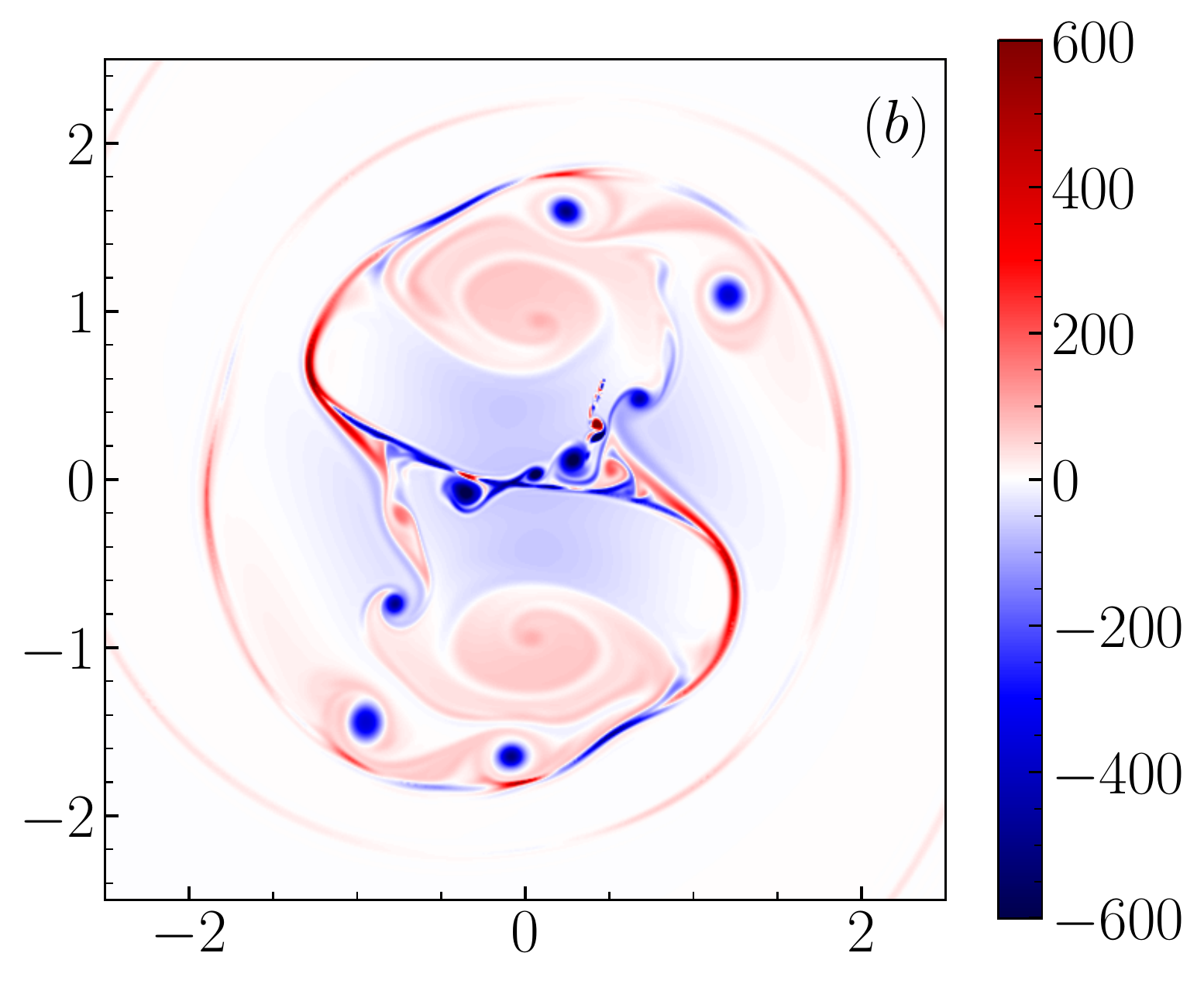}
\includegraphics[scale=0.275]{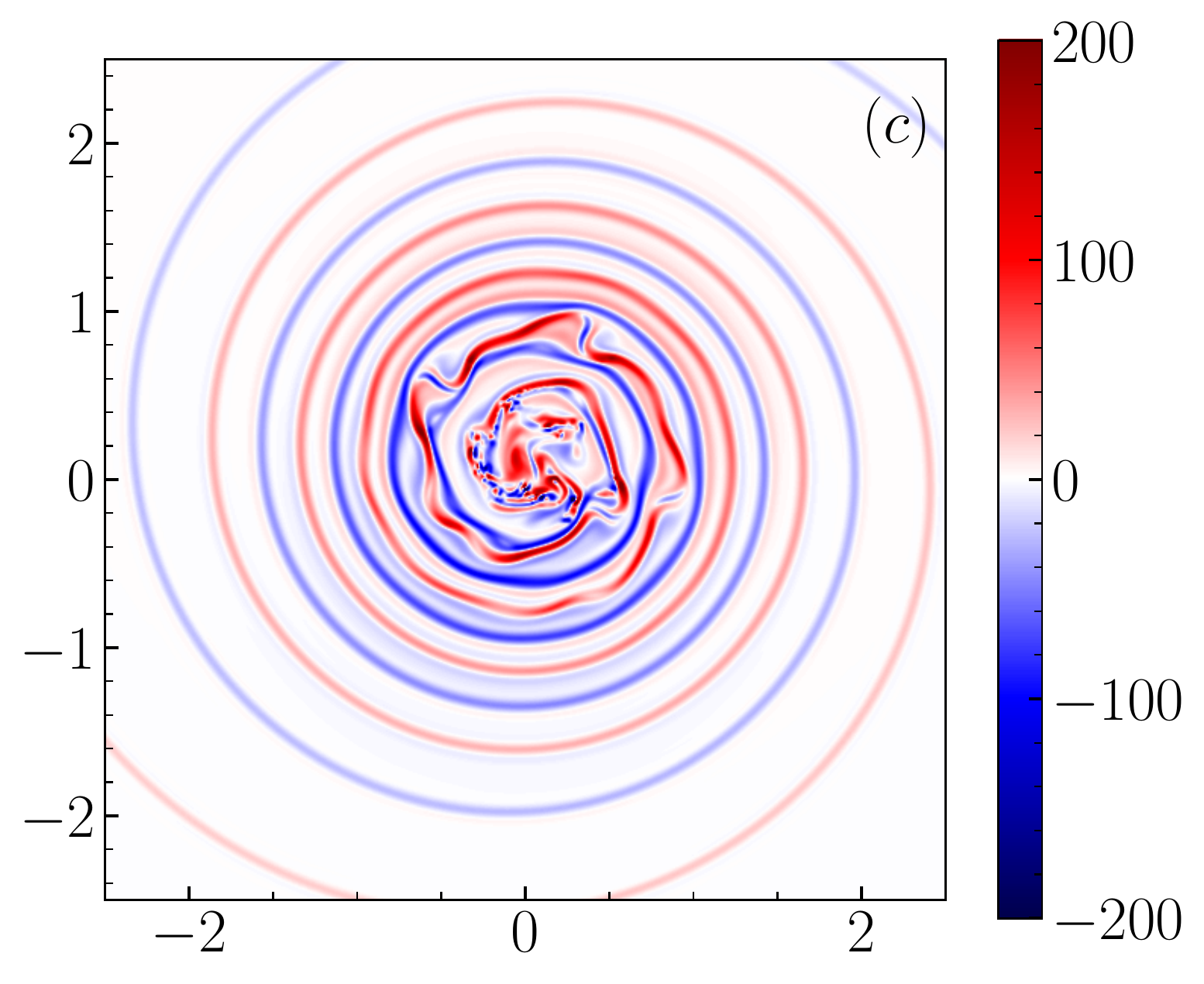}
\caption{Perturbation vorticity  for $We=10040$ and $12$ with and without density contrast. (a) $We=10040$, $At = 0.0$. Perturbation vorticity of small magnitude,  in the neighbourhood of $r \sim r_c$, is generated. Vorticity generated within the core is small. (b) $We=12$, $At = 0$ at $t_n = 25$. (c) $We=10040$, $At = 0.05$ at $t_n = 35$.  Here for both $At$, there are positive and negative vorticity. For $At =0$, the negative vorticity is mainly from the interior of the vortex. }
\label{Fig:vort_contour2}
\end{center}
\end{figure}

The  perturbation vorticity for $We=10040$ and $12$ is plotted in Fig. \ref{Fig:vort_contour2}.  In Fig. \ref{Fig:vort_contour2}(a) where $At = 0$, we find, in accordance with our expectations from the discussion in sections \ref{Sec:vorticity} and \ref{Sec:vorticity_DNS}, that a minuscule amount of vorticity is generated on the interface but no instability happens.  Further, at $r \sim r_c$, this vorticity is positive. However, for $We=12$, with $At = 0.0$, strong negative vorticity is generated within the core, and significant positive vorticity occurs in the vicinity of $r \sim r_c$. As time progresses, vorticity on the interface within the core causes a KH instability, and into a breakdown into small patches of negative vorticity within the core. At even later times, as is evident from Fig. \ref{Fig:vort_contour2}(b)  positive and negative perturbation vorticities are interspersed, and a chaotic state in the core region ensues. This state causes the destruction of the initial Lamb-Oseen vortex. We notice a small level of asymmetry in the vorticity contours shown in \ref{Fig:vort_contour2}(b), and believe this to be a numerical artefact. In a viscous flow, such dynamics would speed up the dissipation of kinetic energy, and also enstrophy in the system, to bring it to a static state. We now examine, in Fig. \ref{Fig:vort_contour2}(c), what happens when there is a density difference in the two fluids. It is clear that when the surface tension is low and the density contrast is high, there is a generation of alternating spirals of positive and negative vorticity in the region outside the core. The blue arms of the spiral are unstable to centrifugal Rayleigh Taylor (CRT) modes, while the red arms are stable \citep{dixit2010}. At later time we see a nonlinear breakdown of the flow, as in the figure. When surface tension and density contrast are both high (see movie3 in the supplementary material), just outside the vortex core, the CRT mode is stabilised by surface tension but displayed at long distances from the core, as is to be expected from our simplified theory. And the vortex gets disrupted due to the rapidly growing KH instability. We observe that the heavy fluid moves outward from the core of the vortex due to centrifugal effects. 

\begin{figure}
\begin{center}
\includegraphics[scale=0.32]{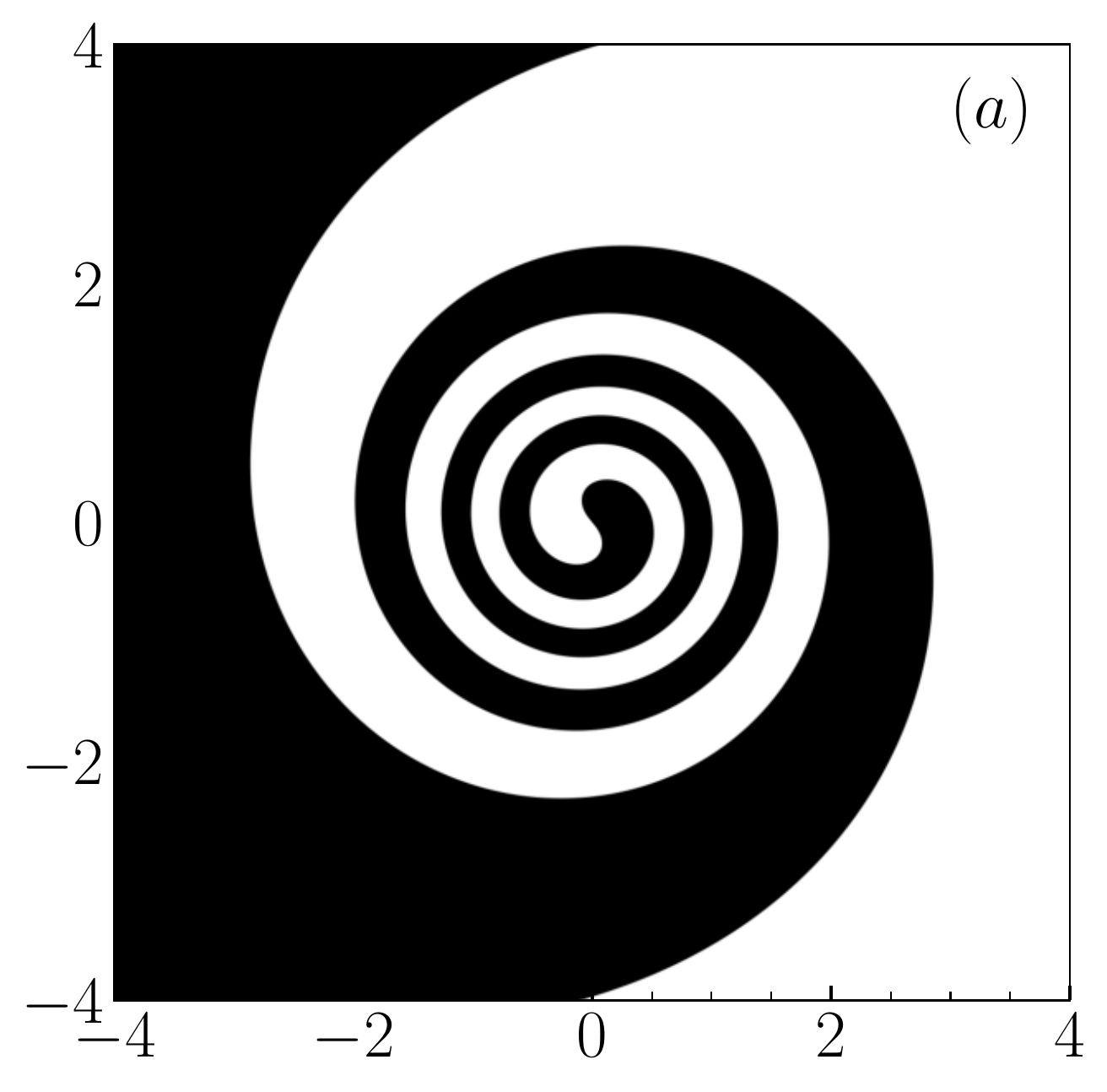}
\includegraphics[scale=0.32]{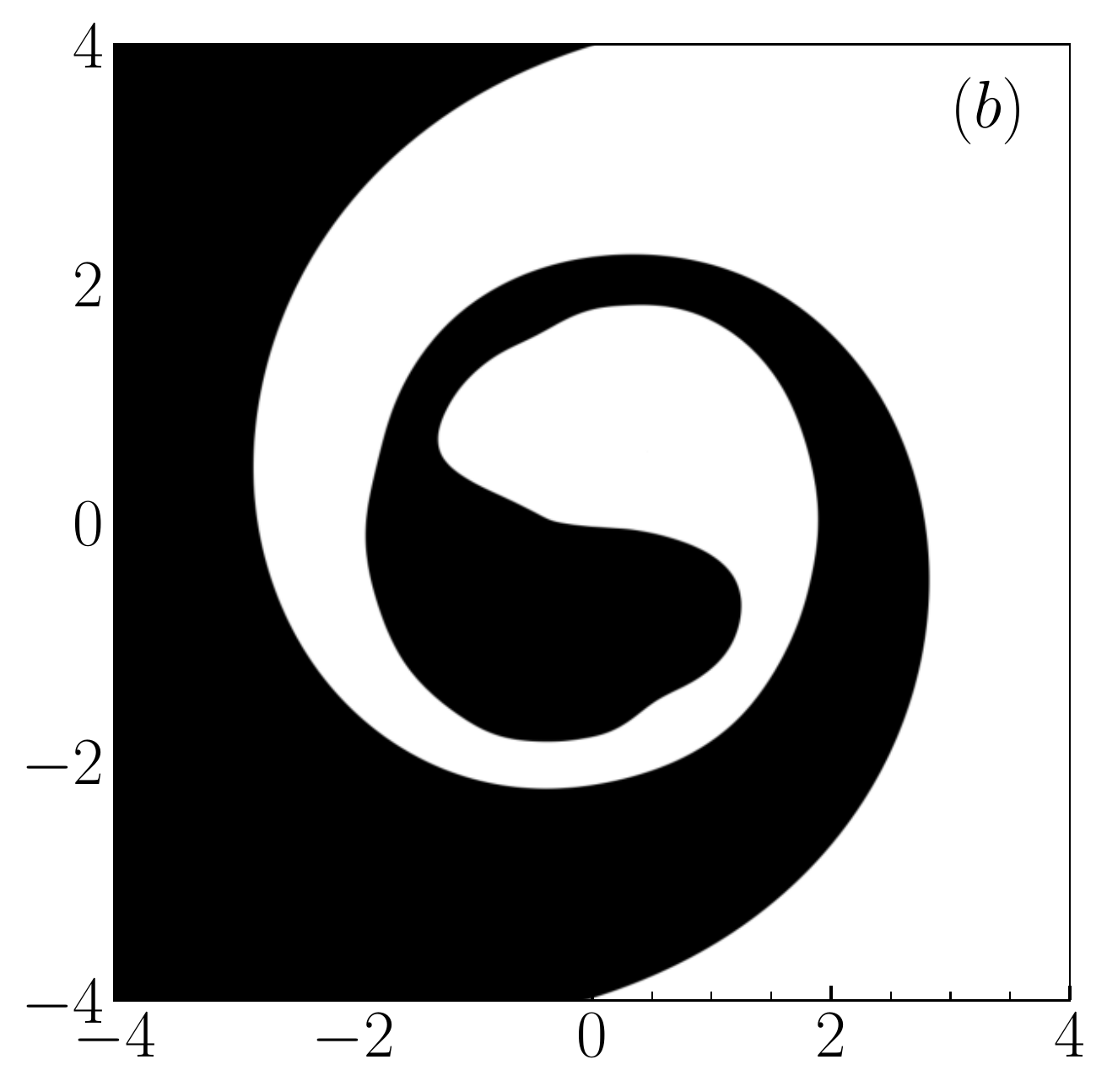}
\includegraphics[scale=0.32]{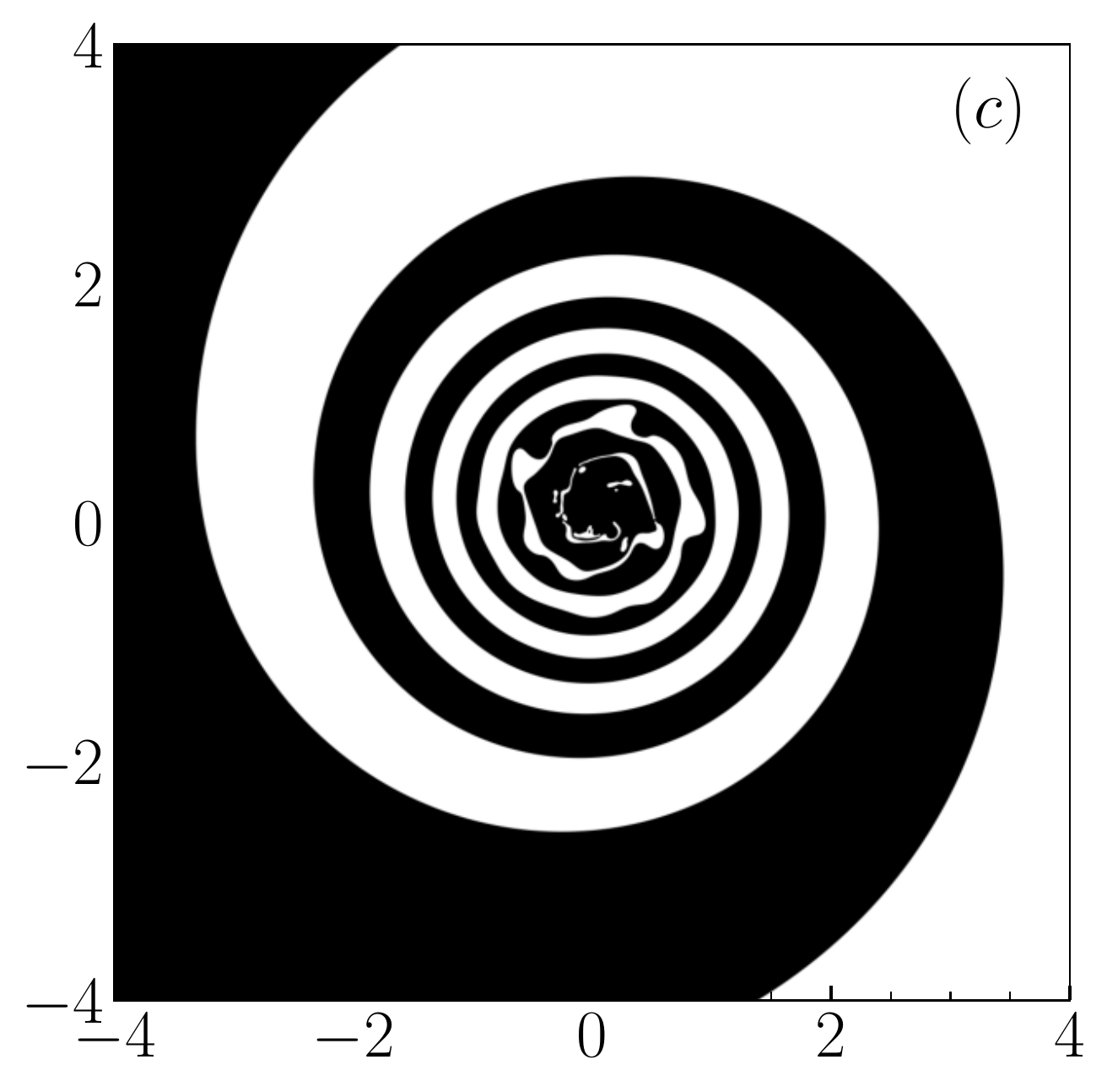}
\caption{Regime occupied by the two fluids, one shown in black and the other in white. (a) A spiral interface is seen at $We=10040$ with $At = 0$ at $t_n=25$. (b) $We=12$, $At=0$, $t_n=25$. The fine structure is erased, and a straight interface is seen in the central portion. (c) $We=10040$ with $At = 0.05$ at $t_n=35$. The CRT instability in the nonlinear regime is visible.}
\label{Fig:density_nocontrast}
\end{center}
\end{figure}

 \subsection{Evolution of the interface}\label{Sec:interface_DNS}

The interfaces for the cases in Fig. \ref{Fig:vort_contour2} are shown in 
Fig. \ref{Fig:density_nocontrast}. At low surface tension and zero density difference, the spiralling interface of Fig. \ref{Fig:density_nocontrast}(a) is indistinguishable from that at zero surface tension, seen earlier in Fig. \ref{Fig:density_profile}(c). Fig. \ref{Fig:density_nocontrast}(b) in combination with Fig. \ref{Fig:vort_contour2}(b) shows that at high surface tension, the interface shape is  completely different from the vorticity distribution. Although vorticity is generated only at the interface, it rolls up into small-scale structures independent of the interface due to the KH instability, and with time is spread through the vortex core. The interface meanwhile adopts a relatively short and straight shape in the central region, in deference to the high surface tension. In contrast, due to the low surface tension in Fig. \ref{Fig:density_nocontrast}(c), the interface closely mimics vorticity contours of \ref{Fig:vort_contour2}(c). Here the CRT instability is on display due to the density contrast.

We thus have a competition between the response to the vortex, which increases the length of the interface, and surface tension, which acts to reduce it. An independent estimate of the interface length can be obtained from energy balance. As there is no external forcing and viscous dissipation here, the total kinetic energy is given by
 \begin{equation}
\partial_t E= \langle \mathbf{u} \cdot \mathbf{F}_{\sigma} \rangle.
 \label{tot_ke}
 \end{equation}
 where the angle brackets refer to an average per unit area taken over the whole domain, and $E =\langle \rho u^2/2 \rangle $.
We also have  $\langle \mathbf{u} \cdot \mathbf{F}_{\sigma} \rangle = - \frac{1}{A}\partial_t \int{\sigma dl}$ where $dl$ is an interfacial line element and $A$ is the total area of the domain \citep{joseph2013stability}, giving
   \begin{equation}\label{balance}
\partial_t  \left(E + \frac{\sigma S}{A} \right) = 0, \quad {\rm i.e.,} \quad - \frac{E-E_0}{ \sigma} =  \frac{S}{A},
\end{equation}
 where $S = \int{dl}$, and $E_0$ the initial kinetic energy. 
 We obtain the length $S$ of the interface numerically as a function of time from all four simulations, by tracking jumps in $c$ from $0$ to $1$ and obtain excellent agreement with Eq.\ref{balance} (not shown). 

\subsection{Two-dimensional turbulence}

\begin{figure}
\includegraphics[scale=0.3]{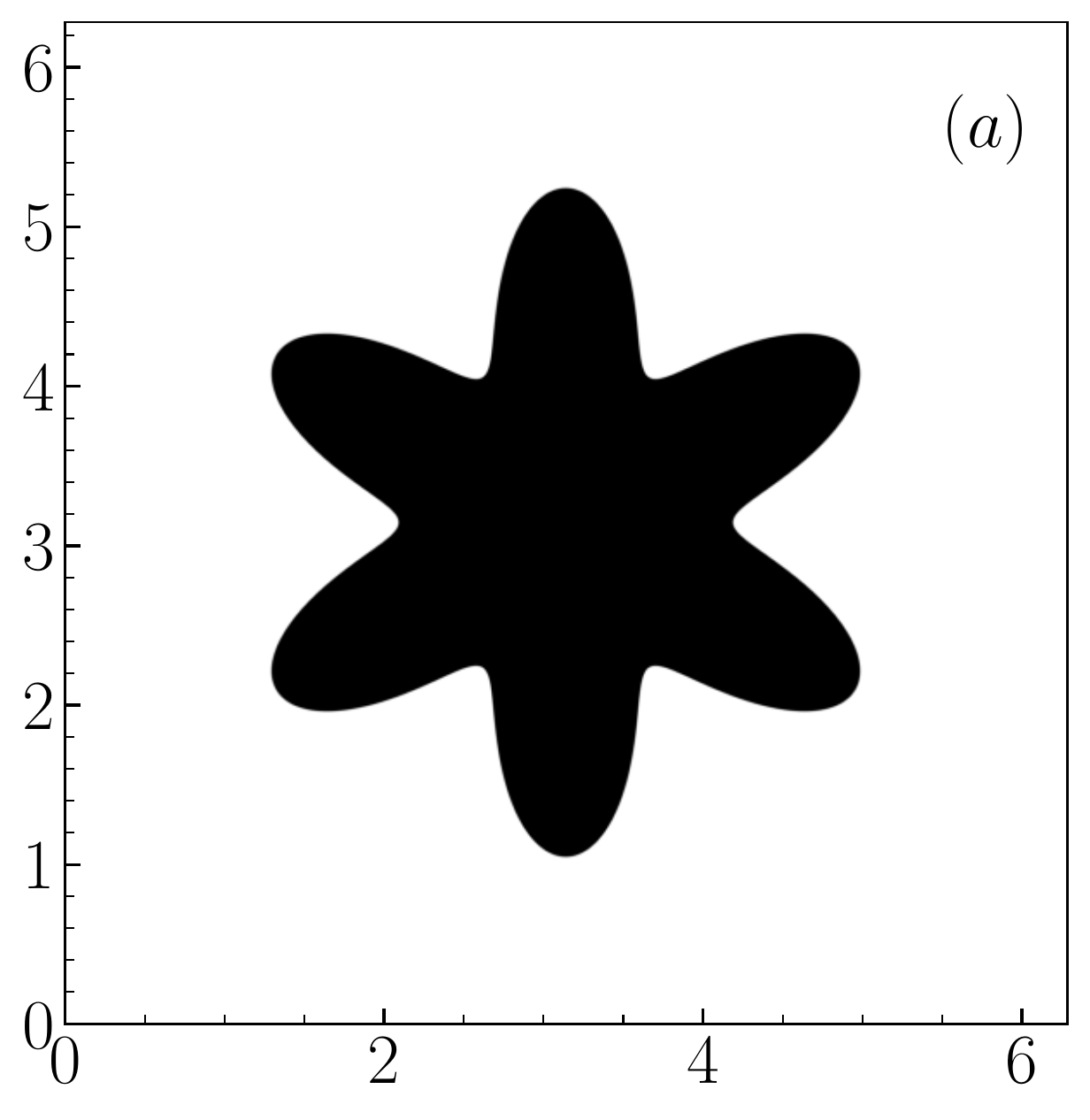}
\hskip 7mm
\includegraphics[scale=0.3]{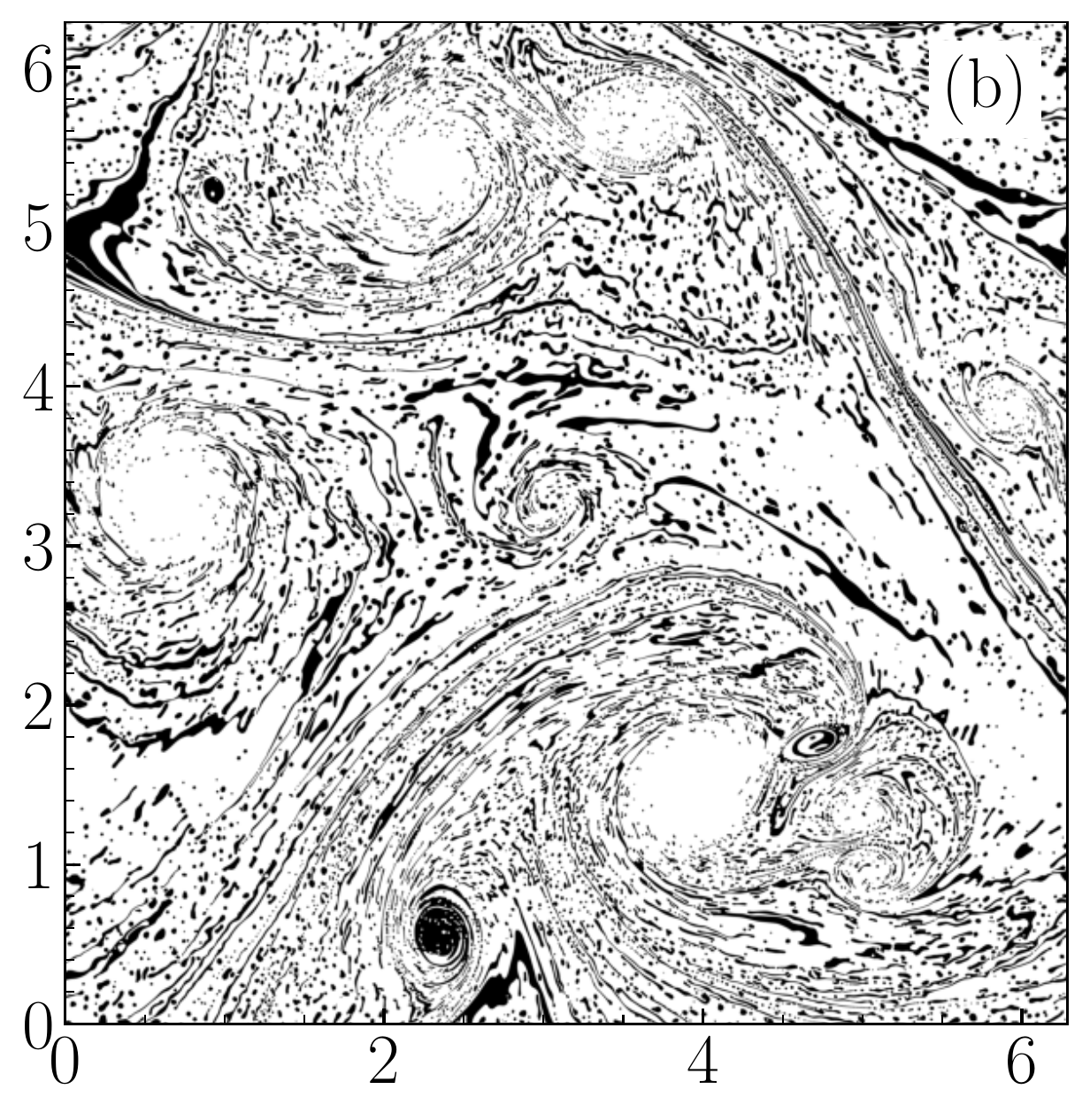}
\hskip 7mm
\includegraphics[scale=0.3]{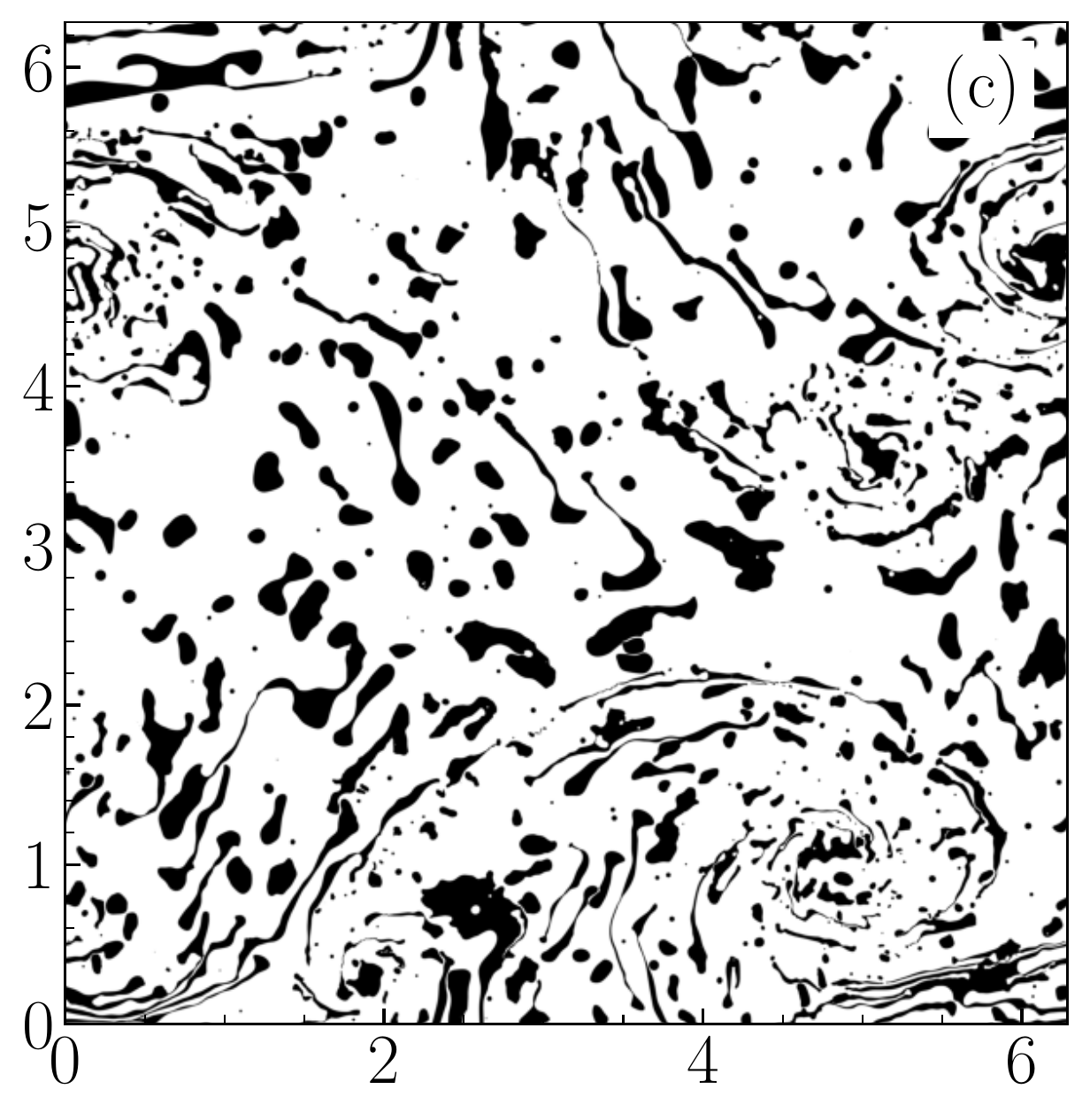}
\\
\includegraphics[scale=0.3]{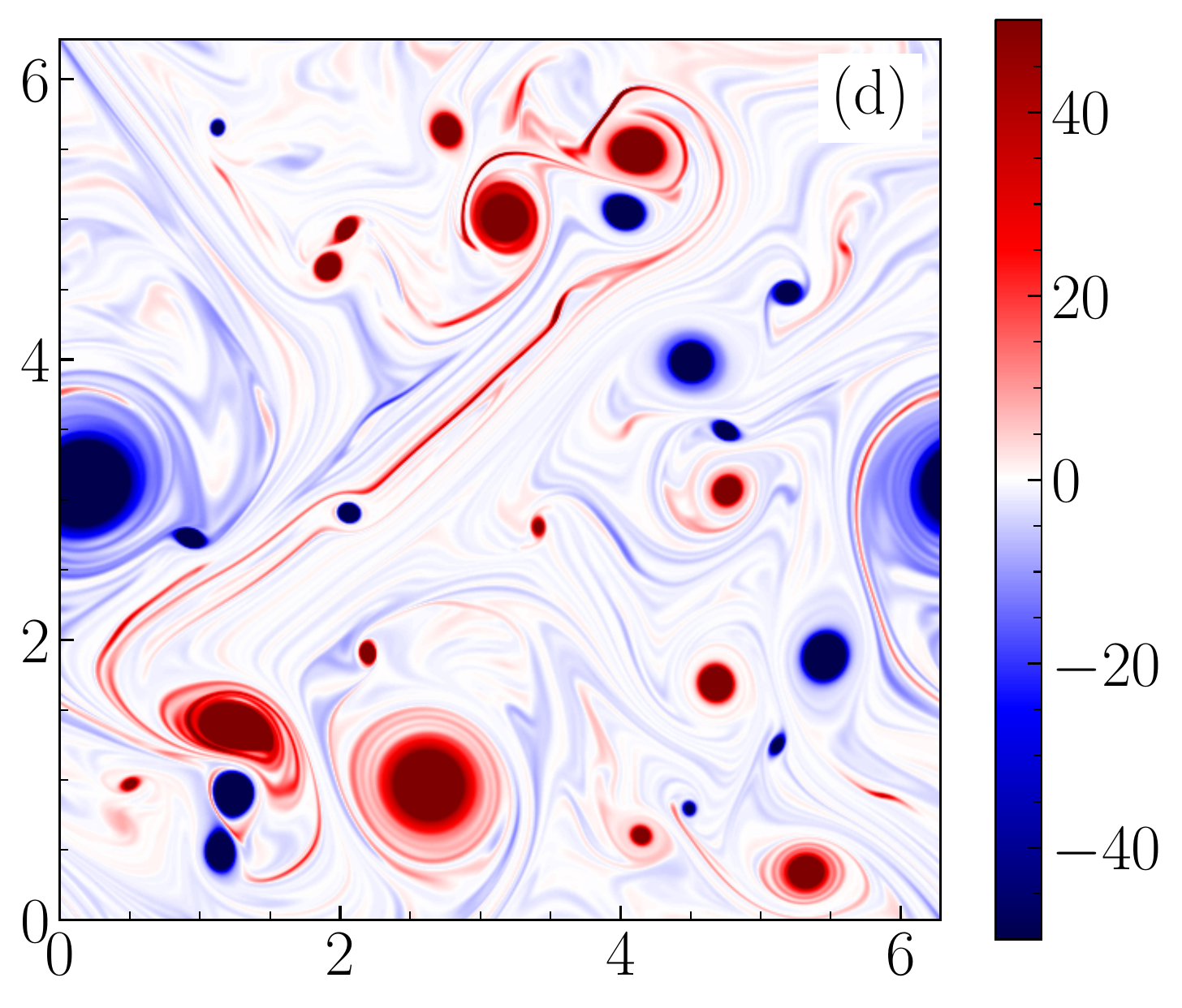}
\includegraphics[scale=0.3]{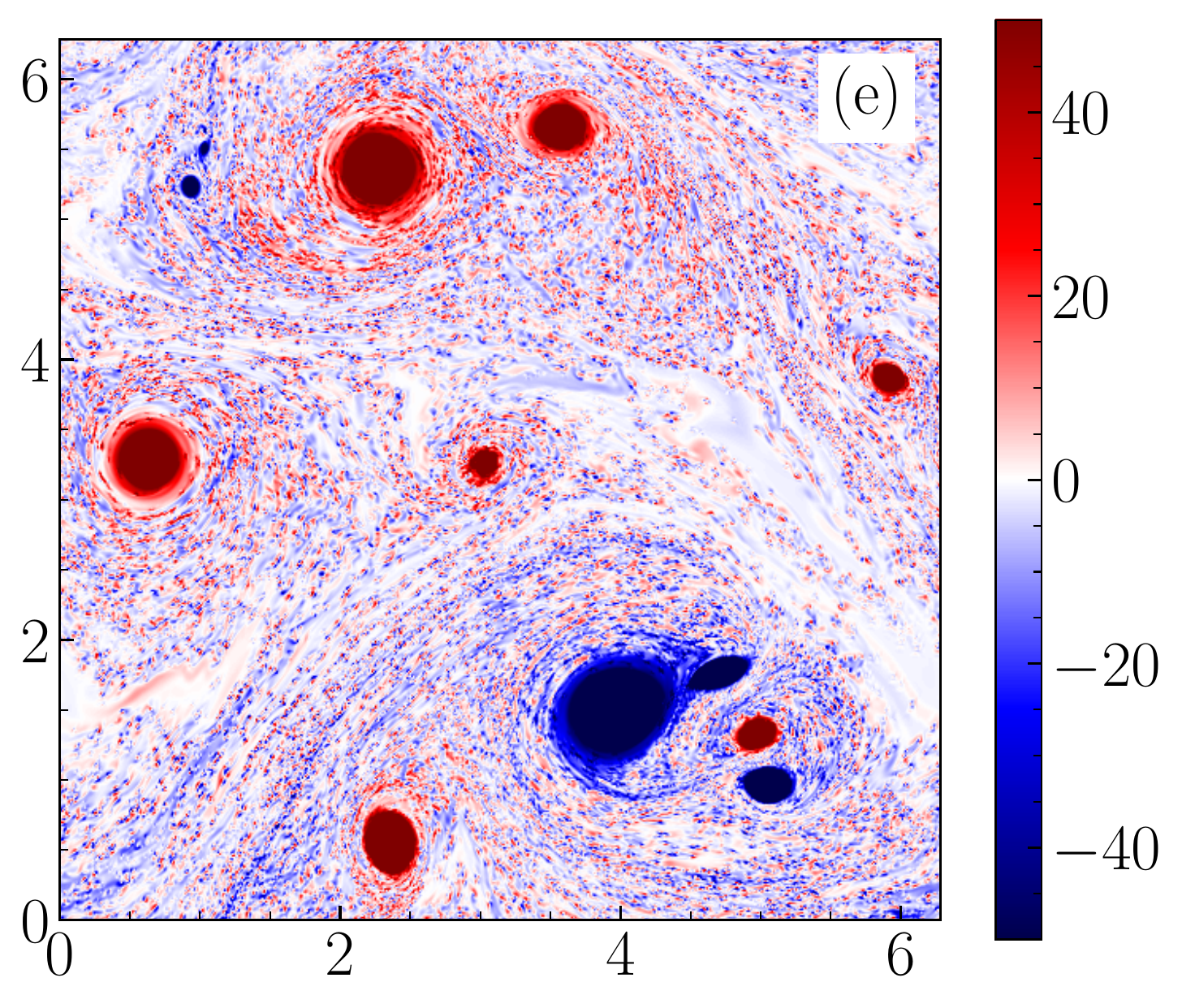}
\includegraphics[scale=0.3]{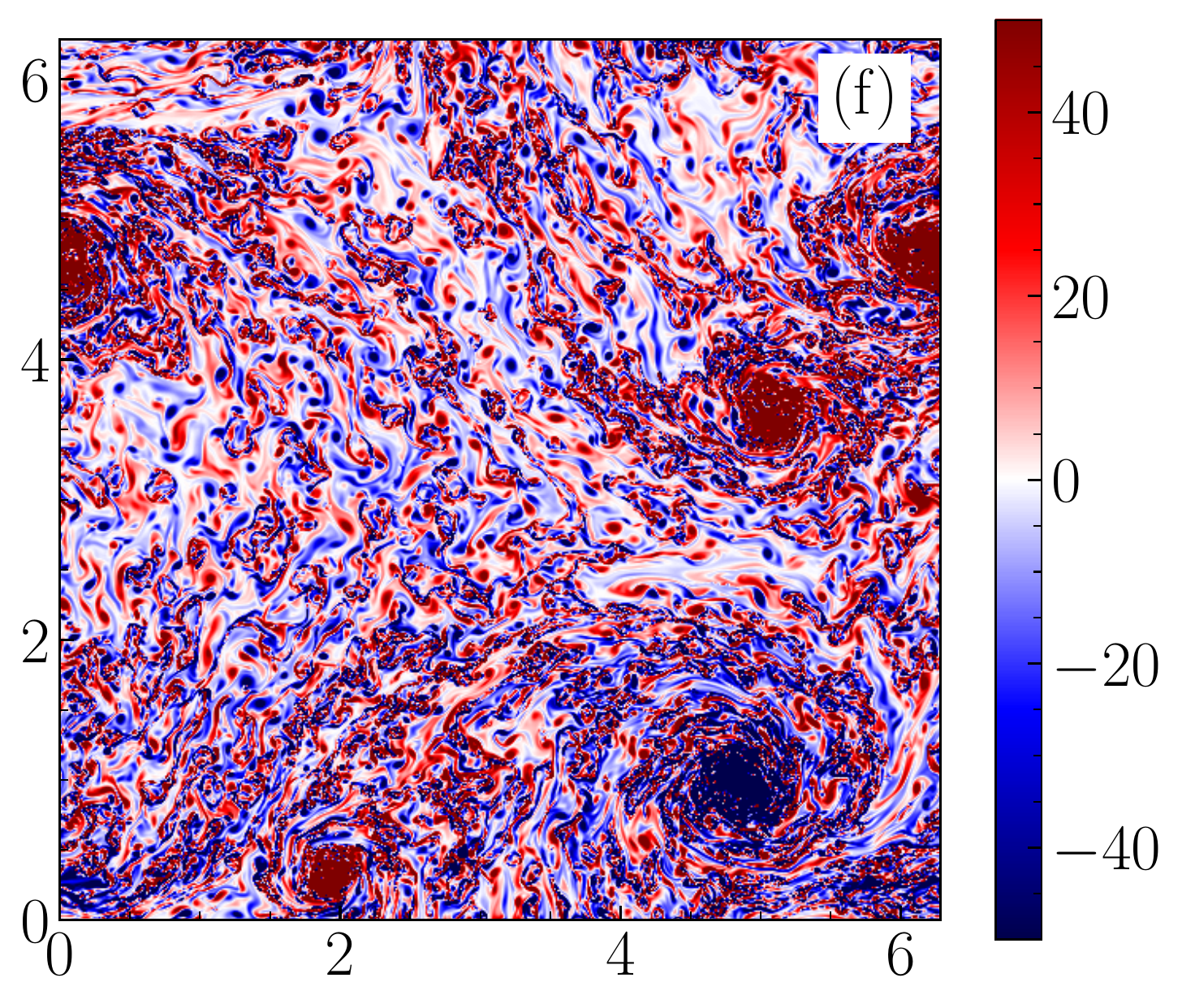}
\caption{Density (a-c) and vorticity (d-f) contours for viscous simulations with $\Rey = 5.3 \times 10^5$. (a,b):  initial profiles, (c,d): profiles at $t/\tau_e = 5.4$ for $We_T=5\times 10^5$. (e,f): profiles at $t/\tau_e = 5.4$ for $We_T=5\times 10^3$. Small-scale vortical structures occur in greater abundance as the surface  tension increases.}
\label{Fig:mv_vort_den_visc}
\end{figure}

To evaluate whether the mechanism we propose, of destabilisation of a flow by surface tension, is of relevance in a general two-dimensional  turbulent flow of two immiscible fluids, we conduct DNS at $At=0$ on a doubly periodic box of length $L=2\pi$ and discretize it with $2048^2$ collocation points. The two fluids are initially separated by a flower-shaped interface (volume fraction of minority phase is $0.225$), which is artificially placed in a turbulent field at zero time, as shown in Figs. \ref{Fig:mv_vort_den_visc}(a) and (d). We perform inviscid as well as viscous simulations -- at a Reynolds number of $5.3 \times 10^5$ -- and find no qualitative difference during the time of the simulation, though in the viscous case we have slowly decaying turbulence. The turbulent flow Weber number is defined as $We_{T} = \rho u^2_{rms} L/\sigma$ where $u_{rms}$ is the root mean square velocity at time $t=0$, and the eddy turn over time is $\tau_e = L/u_{rms}$. \REM{We conduct DNS with  We run the simulations at $2048 \times 2048$ resolution with the volume fraction of black fluid as $0.225$ for four different $We_T$: $\infty$, $5 \times 10^5$, $5 \times 10^4$ and $5 \times 10^3$ in runs $TR1$, $TR2$, $TR3$ and $TR4$ respectively.}

The flow consists of vortices of several scales, encompassing a range of Weber numbers based on each, with the largest being two orders of magnitude smaller than $We_T$. Moreover, the interface is not particularly designed to pass  through the centre of any vortex. So, we do not expect an exact agreement  with theory, but do expect vorticity generation due to surface tension, and instabilities resulting in small-scale vortices. The regions occupied by the two fluids at $t/\tau_e=5.4$ for high and low Weber number are shown in Figs. \ref{Fig:mv_vort_den_visc} (b) and (c) respectively, and the corresponding vorticity distributions are shown in Figs. \ref{Fig:mv_vort_den_visc}(e) and (f) respectively. At $We_T=5 \times 10^5$, the two fluids, which displayed many spiralling interfaces at short times (see movies in the supplementary material), are mingled intricately by this time, with elongated fine structures of the inner fluid. While the cores of the large vortices are still preserved, the smaller vortices undergo mergers and annihilation due to numerical viscosity. This picture is already qualitatively different from flow at zero surface tension, in that we see vorticity generation due to surface tension along the interfaces, as predicted. But the instability, and the vorticity being peeled off from the interfaces is not yet visible. The picture is starkly different at high surface tension. There is an explosion of small-scale vorticity everywhere in the flow, and the initial vortices have been disrupted completely. Because of these small-scale vortices, there is an enhancement of energy at large wave numbers with increase in surface tension. This is consistent with earlier numerical findings \citep{li2009turbulence,trontin2010direct}, which do not give the underlying instability mechanism. We have checked at early times that instability develops as expected.

  \section{Conclusion}
A new role for surface tension, as a destabiliser, in the vortical flow of immiscible fluids is shown here. Due to surface tension, vorticity is generated practically everywhere on the interface, and this vorticity increases with time. 
It follows that the two fluid layers on either side move with different velocities, making it conducive for the KH instability to manifest itself beyond a critical time proportional to $We^{1/4}$. This mechanism acts alongside the CRT instability when there are density differences. Our inviscid simulations on a single vortex confirm our theoretical predictions and also reveal a peeling off of small-scale vorticity from the interface. This mechanism is shown to have a significant presence in viscous and inviscid simulations of two-dimensional turbulence at low Weber number, with an increased proportion of energy in small-scales.

{\bf Acknowledgements} RG acknowledge support of the Department of Atomic Energy, Government of India, under project no. RTI4001.  PP and RR acknowledge support of the Department of Atomic Energy, Government of India, under project no. RTI4007.

\bibliographystyle{jfm.bst}
 \bibliography{references.bib}

  \end{document}